\input{aipcheck}

\documentclass[
    ,final            
  ]
  {aipproc}

\usepackage{graphicx}
\usepackage{color}
\layoutstyle{8x11double}

\newcommand{\bv}[1]{{\bf #1}}

\begin{document}

\title{Constitutive Relations of Jammed Frictionless Granular Materials
under Oscillatory Shear}


\author{Michio Otsuki}{
  address={Department of Physics and Mathematics, Aoyama Gakuin University, Sagamihara 229-8558, Japan}
}

\author{Hisao Hayakawa}{
  address={Yukawa Institute for Theoretical Physics, Kyoto University,
  Kyoto 606-8502, Japan}
}

\begin{abstract}
We numerically investigate the rheological properties 
of jammed frictionless granular materials under an oscillatory shear in terms of our simulation of the distinct element method. 
It is demonstrated  that the constitutive relation 
between the shear stress and 
the shear strain of grains strongly depends on
the amplitude and the frequency of the oscillatory shear.
For a small amplitude region, the granular material behaves as 
a visco-elastic material characterized by the Kelvin-Voigt model,
while it behaves as a {\it yield stress liquid}
for a large amplitude region.
We find that the rheology of grains  
under the large amplitude oscillatory shear
can be described by a phenomenological constitutive model.
\end{abstract}

\maketitle


\section{Introduction}

Athermal disordered materials such as colloidal suspensions, foams,
and granular materials behave as dense liquids when the density is
lower than a critical value, while they behave as amorphous solids
when the density exceeds the critical value \cite{Pusey,Durian,Jaeger}.
This rigidity transition
is known as the jamming transition.

The jamming transition has attracted many physicists since Liu and Nagel indicated its similarity to the glass transition \cite{Liu}.
Recently, the rheology of jammed materials under steady shear has been
extensively studied.
It is remarkable that there exist critical scaling laws 
for the rheological transition
of frictionless grains for the pressure, the elastic modulus, 
and the shear stress,
similar to those in continuous phase transitions,
though the number of contact points is discontinuously 
changed at the jamming point
\cite{OHern02, OHern03, Majmudar,Olsson,Hatano07,Hatano08,Tighe,Hatano10,Otsuki08,Otsuki09,Otsuki10,Nordstrom,Olsson11,Vagberg,Otsuki12,Ikeda,Olsson12}.
This mixed phase transition becomes 
the discontinuous transition associated with
the hysteresis loop 
for frictional granular materials \cite{Otsuki11,bob_nature,Chialvo}.

On the other hand, there exist only a few studies on the rheological properties of jammed materials
near the jamming transition point under an oscillatory shear.
Recently, a numerical simulation of grains in a connected spring network suggested that
there is a critical scaling law for
the complex shear modulus
under an oscillatory shear \cite{Tighe11}.
However, the model in Ref. \cite{Tighe11} underestimates roles of rattlers or the change of connecting networks, and thus, it may be unsuitable to describe realistic jammed materials.
Moreover, as long as we know,
nobody has systematically analyzed rheology of granular materials under the oscillatory shear.

In this paper, we numerically investigate the rheological properties of
jammed frictionless granular materials under the oscillatory shear.
In the next section, we explain our setup and models.
We demonstrate that the relation between the shear stress
and the shear strain strongly depends on the amplitude
and the frequency of the oscillatory shear,
and the rheological properties are consistent with predictions by
a phenomenological constitutive model in the third section.
Finally, we discuss and conclude our results in the last section.

\section{Model and Setup}

Let us simulate a two-dimensional frictionless granular assembly
in a square box with side length $L$ in terms of the distinct element method (DEM).
The system includes $N$ grains, each having
an identical mass $m$. The position and velocity
of a grain $i$ are respectively denoted by
$\bv{r}_i$ and $\bv{v}_i$.
Our system consists of grains having 
the diameters $0.7 \sigma_0$, $0.8 \sigma_0$, $0.9 \sigma_0$, 
and $\sigma_0$, where the number of each species of grains is $N/4$.
The contact force between grains $i$ and $j$ consists of
the elastic part  $\bv{f}_{ij}^{\rm (el)}$
and the dissipative part  $\bv{f}_{ij}^{\rm (dis)}$,
which are respectively given by
\begin{eqnarray}
\bv{f}_{ij}^{\rm (el)} &= &k (\sigma_{ij} - r_{ij})
\Theta (\sigma_{ij} - r_{ij}) \bv{n}_{ij}, \label{Fel}\\
\bv{f}_{ij}^{\rm (dis)} &= &-\eta v_{ij} 
\Theta (\sigma_{ij} - r_{ij}) \bv{n}_{ij}, \label{Fdis}
\end{eqnarray}
where $\bv{n}_{ij}$ is 
$\bv{n}_{ij} = \bv{r}_{ij}/|\bv{r}_{ij}|$ 
with the normal elastic constant $k$,
the viscous constant $\eta$, 
the diameter $\sigma_i$ of grain $i$,
$\bv{r}_{ij} \equiv \bv{r}_{i} - \bv{r}_{j} $, 
$\sigma_{ij} \equiv (\sigma_i + \sigma_j)/2$, and 
$v_{ij} \equiv (\bv{v}_{i}- \bv{v}_{j}) \cdot \bv{n}_{ij}$. 
Here, $\Theta(x)$ is the Heaviside step function characterized by $\Theta(x)=1$
for $x \ge 0$ and $\Theta(x)=0$ for otherwise.

In this paper, we apply an oscillatory shear  
along the  $y$ direction under the Lees-Edwards boundary condition \cite{Evans}.
As a result, there exists macroscopic displacement 
only along the $x$.
The time evolution of such a system, known as the SLLOD system, is given by
\begin{eqnarray}
\frac{d \bv{r}_i}{dt} & = & \frac{\bv{p}_i}{m} + \dot \gamma (t) y_i \bv{e}_x,
\label{SLLOD:1} \\
\frac{d \bv{p}_i}{dt} & = & \sum_{j \neq i} \{ \bv{f}^{\rm (el)}_{ij}+  \bv{f}^{\rm (dis)}_{ij} \}- 
\dot \gamma (t) p_{i,y} \bv{e}_x
\label{SLLOD:2}
\end{eqnarray}
with the peculiar momentum $\bv{p}_i$
and the unit vector parallel to the $x$-direction $\bv{e}_x$.
The shear rate and the shear strain are, respectively, given by
\begin{eqnarray}
\dot \gamma (t) = \dot \gamma_0 \omega \sin(\omega t), \\
\gamma (t) = \dot \gamma_0 \left \{ 1 - \cos(\omega t) \right \}
\end{eqnarray}
with the amplitude $\gamma_0$ and the angular frequency $\omega$.
In this paper, we investigate the shear stress $S$: 
\begin{equation}
S  =   -\frac{1}{L^2}\left <    \sum_i^N \sum_{j>i} 
r_{ij,x} (f_{ij,y}^{\rm (el)} +f_{ij,y}^{\rm (dis)} ) 
\right > 
-\frac{1}{L^2} \left <   \sum_{i=1}^N \frac{p_{x,i}p_{y,i}}{2m} \right >
\label{S:calc}, 
\end{equation}
where 
$\left < \cdot \right >$ represents the ensemble average.

In our simulation $m$, $\sigma_0$, and $k$ are set to be unity, 
and all quantities are converted to dimensionless forms,
where the unit of time scale is $\sqrt{m / k}$.
We use the viscous constant $\eta = 1.0$
This situation corresponds to the constant restitution coefficient
$e = 0.043$. We use the leapfrog algorithm, which is second-order
accurate in time, by using the time interval $\Delta t = 0.2$.
The number $N$ of the particles is $4000$.
We fix the volume fraction of the system $\phi = 0.67$,
which is larger than the critical fraction $\phi_J = 0.645$ 
under a steady shear \cite{Otsuki12}.

\section{RESULTS}
\label{Result:sec}

\begin{figure}
  \includegraphics[height=15em]{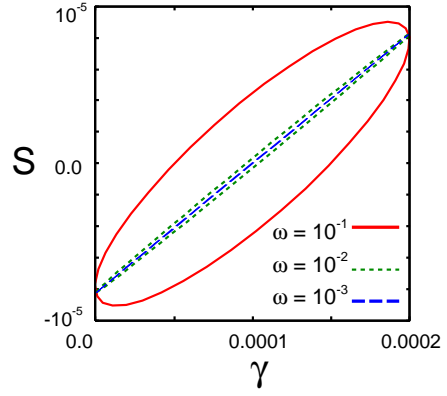}
  \caption{Shear stress $S(t)$ as a function of shear strain $\gamma (t)$
  for the amplitude $\gamma_0 = 10^{-4}$ with $\omega = 10^{-1}, 10^{-2},$ and $10^{-3}$.}
  \label{Sg:4}
\end{figure}

Figure \ref{Sg:4} shows the shear stress $S(t)$ 
as a function of the shear strain $\gamma (t)$
for the amplitude $\gamma_0 = 10^{-4}$ with 
the frequency $\omega = 10^{-1},  10^{-2},$ and $10^{-3}$.
For $\omega = 10^{-1}$, the trajectory draws an ellipse.
The width of it decreases as $\omega$ decreases. 
Finally, the trajectory becomes a single straight line for small $\omega$.
This is a typical property of the visco-elastic materials
described by the Kelvin-Voigt model,
whose constitutive equation is given by
\begin{equation}
S = G \gamma + \eta \dot \gamma
\label{Voigt}
\end{equation}
with the shear modulus $G$ and the shear viscosity $\eta$.
The application of the Kelvin-Voigt model 
might be natural because the shear amplitude is so small that
the change of the contact network is negligible. 

\begin{figure}
  \includegraphics[height=15em]{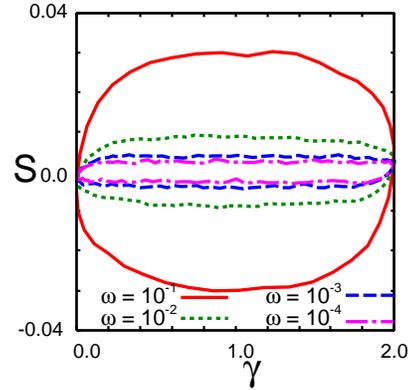}
  \caption{Shear stress $S(t)$ as a function of shear strain $\gamma (t)$
  for the amplitude $\gamma_0 = 10^{0}$ with $\omega = 10^{-1},  10^{-2}, 10^{-3},$ and $10^{-4}$.}
  \label{Sg:0}
\end{figure}

On the other hand, 
the rheology of jammed grains differs from that of the Kelvin-Voigt model
for the larger amplitude $\gamma_0 = 1.0$.
In Fig. \ref{Sg:0}, we plot 
the shear stress $S(t)$ as a function of the shear strain $\gamma (t)$
for the amplitude $\gamma_0 = 10^{0}$ 
with $\omega = 10^{-1}, 10^{-2}, 10^{-3},$ and $10^{-4}$.
As the frequency decreases, the width of the trajectory decreases,
but it still remains even in the smallest $\omega$.
For the smallest frequency $\omega=10^{-4}$,
the shear stress $S$ shows a linear elastic dependence on the shear strain
in the region $0 < \gamma < 0.2$ with $S>0$ and $1.8 < \gamma < 2\gamma_0$
with $S<0$,
while $|S|$ is almost constant for the other region.
The existence of the width in the limit $\omega \to 0$
is a typical property of {\it yield stress fluids} \cite{Ewoldt}.

In order to describe the rheological properties of yield stress fluids,
such as aqueous foams \cite{Rouyer},
the elastic Bingham model has been used \cite{Ewoldt,Yoshimura}.
The shear stress-strain relationships for the elastic Bingham model 
are given by
\begin{eqnarray}
S =\left\{ \begin{array}{ll}
G \gamma_E, & |\gamma_E|<\gamma_c\\
G \gamma_c + \eta \dot \gamma, & |\gamma_E|>\gamma_c,
\end{array} \right.
\end{eqnarray}
where $\gamma_c$ is the yield strain and $\gamma_E$ is the recoverable elastic
strain $\gamma_E$. The elastic strain $\gamma_E$ is obtained by integrating
the shear rate with respect to time, but saturates at $|\gamma_E| = \gamma_c$.
When the shear stops and the direction of it is reversed, the accumulated elastic strain
is recovered and the response is linear elastic until the material
is re-yielded again for $|\gamma_E| = \gamma_c$.

In the elastic Bingham model, the shear stress under steady shear
with the shear rate $\dot \gamma$ is given by
\begin{equation}
S = G \gamma_c + \eta \dot \gamma,
\label{Bingham:steady}
\end{equation}
but this is inconsistent with the constitutive equation of
frictionless granular materials under a steady shear.
Indeed, 
the shear stress $S$ of the jammed frictionless granular materials 
under the steady shear satisfies
a critical scaling law given by \cite{Otsuki08,Otsuki09}
\begin{equation}
S = (\phi - \phi_J)^{y_\phi} F\left [ \dot \gamma / (\phi - \phi_j)
^{y_\phi/y_\gamma}\right ],
\end{equation}
where the scaling function $F(x)$ obeys
\begin{eqnarray}
\lim_{x\to0} F(x) & = & \mbox{const.}, \\
\lim_{x\to \infty } F(x) & \propto & \dot \gamma ^{y_\gamma}.
\end{eqnarray}
The critical exponents $y_\phi$ and $y_\gamma$
for the interaction force in Eq. (\ref{Fel})
is approximately given by 
\begin{equation}
y_\phi = 1, \qquad y_\gamma = 2 / 5,
\end{equation}
which are also phenomenologically derived \cite{Otsuki08,Otsuki09}.
Hence, 
the shear stress $S$ of the jammed frictionless granular materials 
under steady shear is expected to satisfy
\begin{equation}
S = S_c + \eta' \dot \gamma ^{2/5}
\label{Otsuki:steady}
\end{equation}
with a constant $\eta'$ and the yield stress $S_c$,
which differs from that of the Kelvin-Voigt model (\ref{Bingham:steady}).

In order to characterize the rheology of grains as in Fig. \ref{Sg:0},
we propose a new phenomenological constitutive equation, where the shear stress $S$
is given by
\begin{eqnarray}
S =\left\{ \begin{array}{ll}
G \gamma_E + \eta' \dot \gamma^{2/5}, & |\gamma_E|<\gamma_c\\
G \gamma_c + \eta' \dot \gamma^{2/5}, & |\gamma_E|>\gamma_c.
\end{array} \right.
\label{phenomenological_model}
\end{eqnarray}
Here, we assume that the shear stress consists of
the dynamic part given by $\eta' \dot \gamma^{2/5}$
and the static part, which is proportional to the elastic strain $\gamma_E$ 
until the strain exceeds the yield strain $\gamma_c$,
where the change of the contact network appears.

\begin{figure}
  \includegraphics[height=15em]{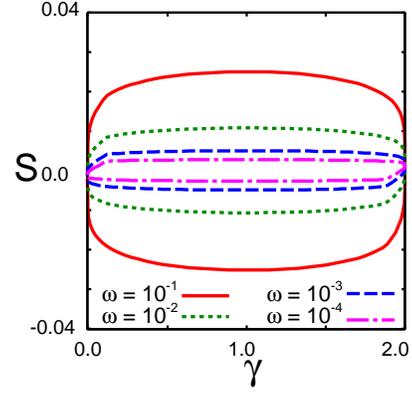}
  \caption{Shear stress $S(t)$ as a function of shear strain $\gamma (t)$
 obtained from the phenomenological model (\ref{phenomenological_model}) for the amplitude $\gamma_0 = 10^{0}$ with $\omega = 10^{-1},  10^{-2},$ and $10^{-3}$.}
 \label{model:fig}
\end{figure}

In Fig. \ref{model:fig}, we plot
the shear stress $S(t)$ as a function of the shear strain $\gamma (t)$
obtained from the phenomenological model (\ref{phenomenological_model}) 
for the amplitude 
$\gamma_0 = 10^{0}$ with $\omega = 10^{-1}, 10^{-2}, 10^{-3},$ and $10^{-4}$.
Here, we choose the parameters as
$G = 0.02$, $\eta' = 0.06$, and $\gamma_c = 0.06$.
From the comparison of Fig. \ref{Sg:0}  with Fig. \ref{model:fig},
we find that our phenomenology given 
by Eq. (\ref{phenomenological_model}) provides semi-quantitatively
accurate behavior of DEM shown in Fig. \ref{Sg:0}.

\section{Discussion and Conclusion}
\label{Discussion:sec}

It is known that 
 the rheological properties of granular materials under a steady shear
 drastically depend on the volume fraction near the jamming transition point.
\cite{Hatano07,Hatano08,Hatano10,Otsuki08,Otsuki09,Otsuki10,Otsuki12}.
We also expect that the rheological properties of 
granular materials under an oscillatory shear strongly depend on the volume
fraction $\phi$.
We will report such a $\phi$-dependence of the constitutive equation elsewhere.

In this paper, we restrict our interest to the frictionless
particles. When the particles have friction,
the rheological properties under oscillatory shear should be changed.
In fact, 
the critical properties and the critical fraction
depend on the friction coefficient for frictional granular particles under steady shear \cite{Otsuki11,Chialvo}.
The rheological properties in the dynamical systems of frictional particles
under oscillatory shear will be discussed in our future work.

When the amplitude $\gamma_0$ becomes small,
the constitutive model given by Eq. (\ref{phenomenological_model})
converges to
\begin{equation}
S = G \gamma + \eta' \dot \gamma ^{2/5},
\end{equation}
which differs from the Kelvin-Voigt model (\ref{Voigt}).
For small $\gamma_0$,
we expect that the rheology is characterized by the Kelvin-Voigt.
Hence, we should improve the constitutive model in order to
describe the rheological properties in wide range of $\gamma_0$.

In conclusion, we have numerically studied frictionless granular materials 
under oscillatory shear, and demonstrated that the rheology
depends on the amplitude of the shear. We also proposed a new phenomenological
constitutive equation, and
demonstrated that the rheology 
for the large amplitude is well described by our phenomenological model.


\begin{theacknowledgments}
We thank K. Miyazaki for valuable discussions. 
This work is partially supported by the Ministry of Education, 
Culture, Science and Technology (MEXT), Japan (Grant Nos. 21540384) 
and the Grant-in-Aid for the global COE program 
``The Next Generation of Physics, Spun from Universality and Emergence'' from MEXT, Japna.
\end{theacknowledgments}



\bibliographystyle{aipproc}   



\end{document}